\documentclass[11pt,twoside]{article}
\usepackage{asp2010}
\aspcpryear{2010}
\aspvoltitle{UP: Have Observations Revealed a Variable Upper End of
  the Initial Mass Function?}

\resetcounters

\begin{document}

\title{Accretion-regulated star formation in late-type galaxies}
\author{Jan Pflamm-Altenburg$^1$ and Gerhard~Hensler$^2$
\affil{$^1$Argelander-Institut f\"ur Astronomie (AIfA), Universit\"at Bonn, 
  Auf dem H\"ugel 71, D-53121 Bonn, Germany}
\affil{$^2$Institut f\"ur Astronomie, Universit\"at Wien, 
  T\"urkenschanzstr.\ 17, A-1180 Wien, Austria}
}

\begin{abstract}
  We develop a four-phase galaxy evolution model in order to study the effect
  of accretion of extra-galactic gas on the star formation rate (SFR) 
  of a galaxy. Pure self-regulated star formation of isolated galaxies is replaced
  by an accretion-regulated star formation mode. The SFR settles into 
  an equlibrium determined entirely by the gas accretion rate on a Gyr 
  time scale. 
  
\end{abstract}

\section{Introduction}
The galactic star formation process is understood to be self-regulated:
An increasing star formation activity leads to an increase of the energetic
feedback to the interstellar medium (ISM) by supernovae, stellar radiation
and stellar winds heating up the cold gas component from which stars form, 
suppressing further star formation. After decrease of stellar feedback 
on-going cooling of the ISM leads to an increase of the cold gas component and
star formation rises again. These negative feedback processes lead to a 
self-regulated equilibrium of star formation \citep*{koeppen1995a}. 
The star formation rate (SFR) is correlated with 
the cold gas density as found by \citet{schmidt1959a} 
or with the gas surface density \citep{kennicutt1998a}. An isolated galaxy  would then be 
depleted in gas and the SFR decreases continuously. 

Indeed large gas reservoirs have been found around star forming galaxies which are
able to refuel galaxies. In individual studies of local star forming 
galaxies gas accretion rates have been estimated by dividing the mass of
extra-galactic gas stored in the extended gas reservoirs through the infall
time scale. These calculated accretion rates are of the order of the current
SFR which has lead to the conclusion that gas accretion can 
sustain long-term high-level SFRs. This requires that 
the SFR settles into an equilibrium determined
by the accretion rate and that the self-regulated mode of 
 galactic star formation
is replaced by an accretion regulated
process.

We here quantify with a galactic four-phase model if and how fast the
accretion regulated star formation equilibrium is established.

\section{Model}

\begin{figure}
  \plotone{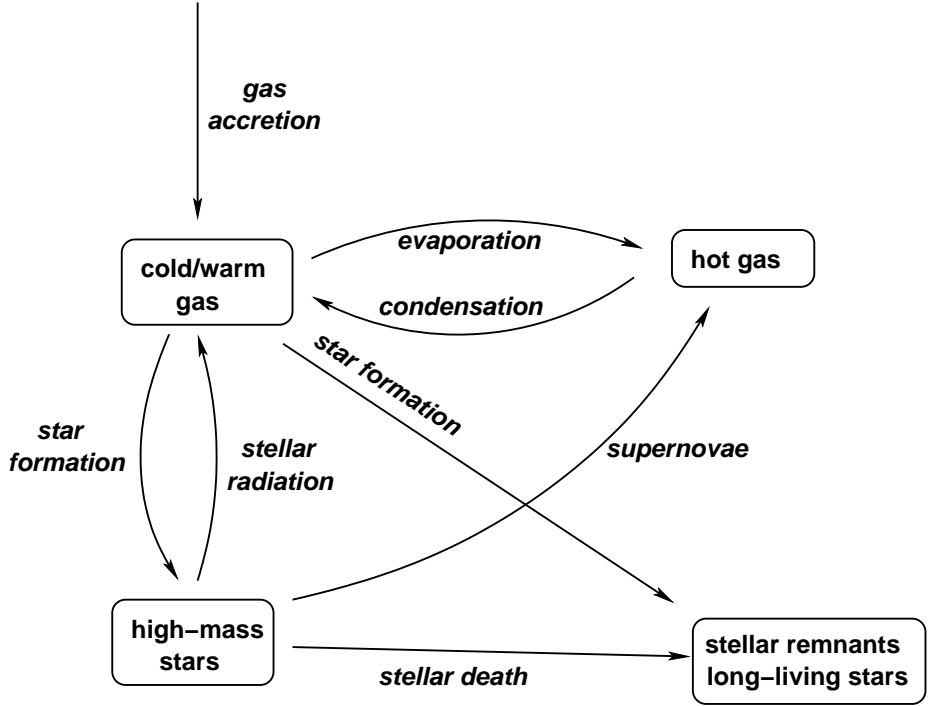}
\caption{\label{fig_cycle} A sketch of the four-phase model 
for galaxy evolution (Pflamm-Altenburg \& Hensler, in prep.) illustrating 
the exchange of matter and energy between the phases.}
\end{figure}

The galaxy evolution model used here to study the effect of gas accretion 
is the four phase model developed in \citet*{koeppen1998a} for an isolated
galaxy without gas accretion. The four phases are: 
i) hot gas, $g$, fed by supernovae, 
ii) cold and warm gas, $c$, from which stars form, 
iii) high-mass stars, $s$, driving the energetic feedback, and 
iv) remnants, $r$, which are long-lived low-mass stars and stellar 
remnants of high-mass stars, i.e. black holes and neutron stars. 
The matter and energy cycle in this four phase model is illustrated
in Figure~\ref{fig_cycle}. 

The galaxy evolution model is described by four equations quantifying the
time derivatives of the mass density of the four phases,
\begin{equation}\label{eq_c}
  \dot c = -\Psi -E_c + K_g+A_c 
\end{equation}
\begin{equation}\label{eq_g}
  \dot g = \frac{\eta}{\tau} s+E_c - K_g
\end{equation}
\begin{equation}\label{eq_s}
  \dot s = \xi\Psi -\frac{1}{\tau}s 
\end{equation}
\begin{equation}\label{eq_r}
  \dot r = (1-\xi)\Psi +\frac{1-\eta}{\tau}s\;,
\end{equation}
where $A_c$ is the accretion rate, $\Psi$ the star formation rate, 
$E_\mathrm{c}$  the 
evaporation rate of cold and warm cloud  material, $K_\mathrm{g}$
the condensation rate of gas, $\xi$ the formation fraction of
high-mass stars, $\eta$ their gas return fraction and $\tau$ their
mean life time.

The exchange of energy between hot gas, $g$, and cold and warm gas, 
$c$, is formulated by 
\begin{equation}\label{eq_eg}
  \dot e_g =  h_\mathrm{SN} s -g^2 \Lambda_0(T_g)+E_c b\bar T_c-K_g b T_g
\end{equation}
\begin{equation}\label{eq_ec}
  \dot e_c =   h_\gamma s -c^2 \Lambda_0(T_c)-E_c b T_c+ K_g b \bar T_g\\
  -\Psi b T_c + b T_{A_c} A_c +\frac{1}{2}v_c^2 A_c\;,
\end{equation}
where $h_\mathrm{SN}$ is the heating coefficient by supernovae, 
$h_\gamma$ the heating coefficient by ionising radiation,
$\Lambda_0$
the cooling function, $T_c$ the temperature of the clouds, 
$\bar T_c$ temperature of evaporated clouds
when entering the hot gas, and $T_g$ the temperature of the hot gas,
$\bar T_g$ the temperature of the hot gas when condensing onto  
the clouds, $T_{A_c}$ is the temperature and $v_c$ the infall velocity
of the accreted gas.

A full description of the calculation of the parameters can be found in 
Pflamm-Altenburg \& Hensler (in prep.).

\section{Constant accretion}

In the case of constant accretion it is expected that the SFR
will settle into an equilibrium. An expression for the
equilibrium can be derived by setting $\dot g=\dot c=\dot s = 0$
in eq.~\ref{eq_c}--\ref{eq_r}. The equilibrium SFR is then determined by the 
accretion rate, 
\begin{equation}
   \label{eq_equi}
  \Psi = \frac{1}{1-\xi\eta}A_c\;.
\end{equation}
For typical values of $\xi\approx 0.9$ and $\eta\approx 0.1$ 
(Pflamm-Altenburg \& Hensler, in prep.) it follows that the accretion rate 
amounts to $\approx$ 90~\% of the equilibrium SFR. 
The remaining 10~\% is covered by returned material from high-mass stars.

Note that the equilibrium SFR is independent on the local formulation of the
star formation law, i.e. the slope of a Schmidt-type or Kennicutt-Schmidt-type
description of the star formation rate density or surface density. This implies
that star formation, which is locally self-regulated and described by a 
gas-density
dependent formula, is globally described  by an accretion regulated mode.

\begin{figure}
  \plottwo{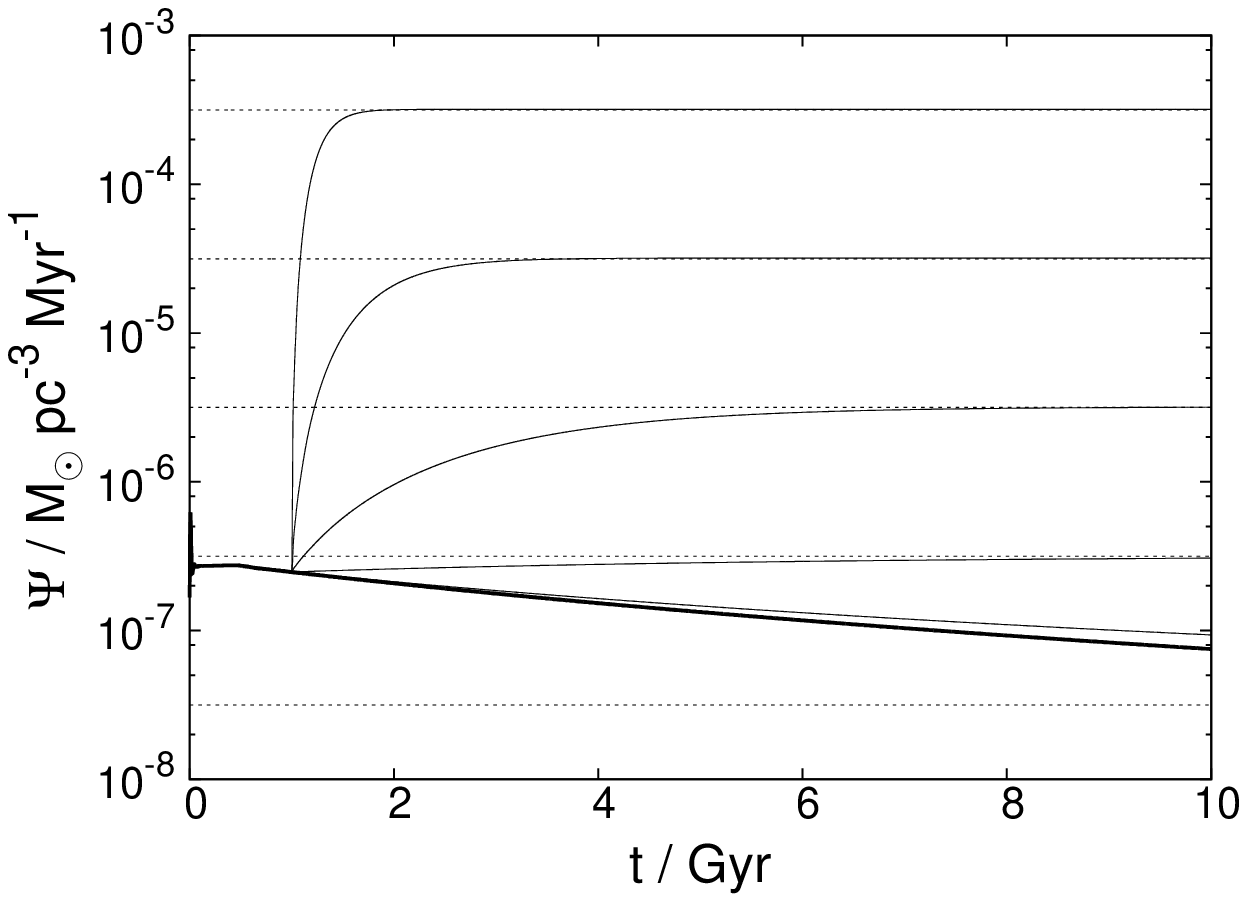}{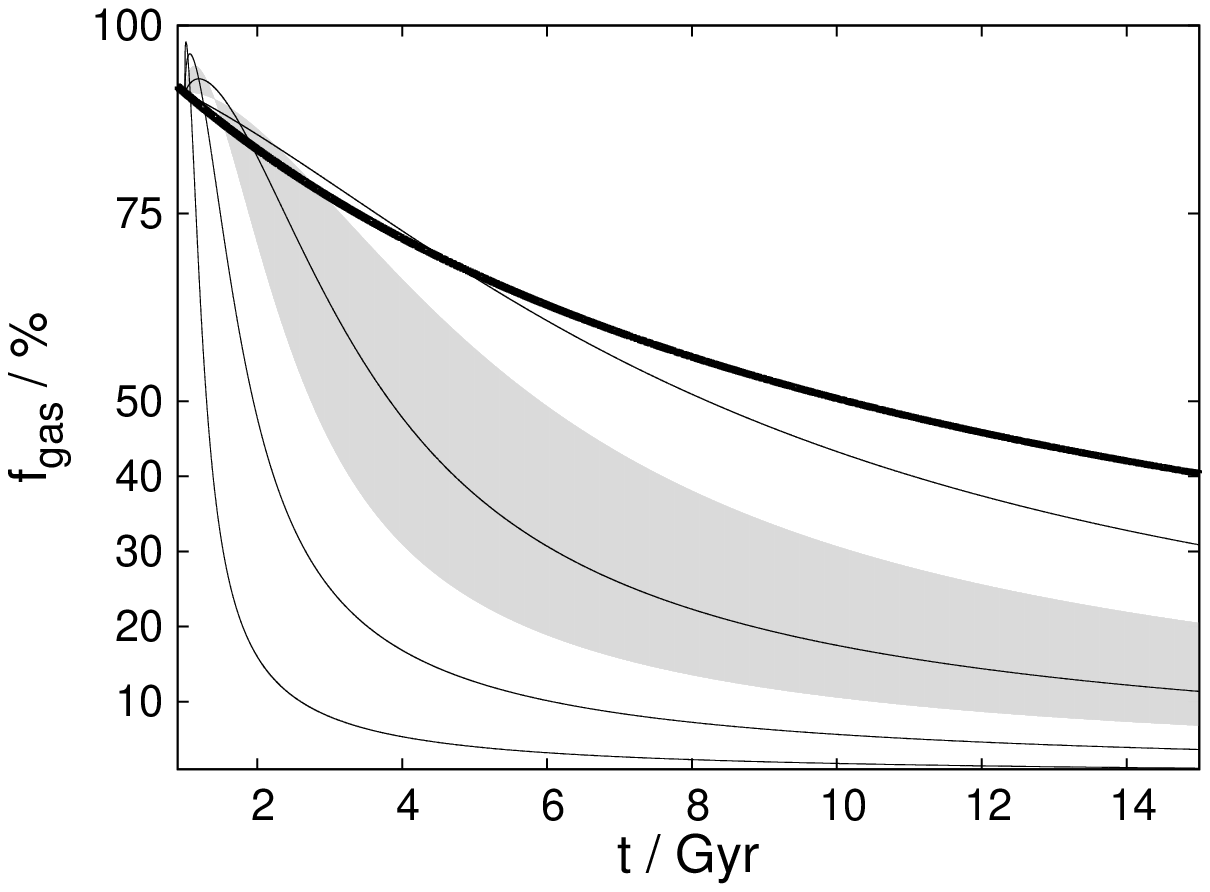}
\caption{\label{fig_sfr_f_gas}Left: The evolution of the SFR density for different accretion rates: $2.83\times 10^{-4}$, $2.83\times 10^{-5}$, $2.83\times 10^{-6}$, $2.83\times 10^{-7}$, and $2.83\times 10^{-8}$~M$_\odot$~pc$^{-3}$~Myr$^{-1}$ (from top to bottom). The thick solid line shows the case of pure 
  self-regulation without accretion. The dashed lines mark the equilibrium
SFRs calculated with eq.~\ref{eq_equi}. 
Right: The evolution of the gas mass fractions for different accretion rates: $2.83\times 10^{-4}$, $2.83\times 10^{-5}$, $2.83\times 10^{-6}$, $2.83\times 10^{-7}$, and $2.83\times 10^{-8}$~M$_\odot$~pc$^{-3}$~Myr$^{-1}$ (from bottom to top).
The solid line represents the result of self-regulated star formation. The grey shaded area shows the region of reasonable accretion rates of Milky-Way-type
galaxies.}
\end{figure}

If and how fast this equilibrium is reached can be studied by
numerical integration of the set of equations \ref{eq_c}--\ref{eq_ec}.
Fig~\ref{fig_sfr_f_gas} (left) shows the evolution of the SFR density 
for different accretion rates (solid curves). 
For comparison, the thick solid line shows the case of pure 
  self-regulation without accretion. The dashed lines mark the equilibrium
SFRs calculated with eq.~\ref{eq_equi}. It can be seen that in the case of 
constant accretion the SFR approaches the equilibrium value on a Gyr time scale
if the accretion rate drives an equilibrium SFR which is larger than the corresponding self-regulation SFR. If the equilibrium SFR determined by the 
accretion rate  is lower than the self-regulation SFR then the evolution follows the self-regulated star-formation mode.

The evolution of the corresponding gas mass fractions are shown in 
Fig.~\ref{fig_sfr_f_gas} (right). It can be seen that the gas mass fractions
decrease faster with increasing accretion rate. An isolated galaxy which is
purely self-regulated consumes the gas on the longest time scale, whereas galaxies 
with highest accretion rates experience high SFRs so that the gas mass fractions becomes the lowest.
 
\section{Conclusion}
Using a four-phase galaxy evolution model we have shown that in the case of 
accretion of extra-galactic gas the global star formation process switches 
from pure self-regulation into an accretion regulated mode. Under these circumstances the accretion-regulated SFR is simply determined by the accretion rate
(eq.~\ref{eq_equi}). If large-scale star formation is accretion-regulated 
then a varying gas accretion history implies naturally a varying star 
formation history. 
Additonally, as star-forming galaxies are extended objects an inhomogeneous
infall is expected to cause an inhomogenous star formation pattern.
These issues, how closely the star formation history is linked to the gas 
accretion history and how an inhomogeneous or radial varying gas infall 
would lead to an inhomogenous or radial varying SFR will
be explored in future studies.

\bibliography{pflamm-altenburg_j}

\begin{thebibliography}{}
\expandafter\ifx\csname natexlab\endcsname\relax\def\natexlab#1{#1}\fi
\expandafter\ifx\csname url\endcsname\relax
  \def\url#1{\texttt{#1}}\fi
\expandafter\ifx\csname urlprefix\endcsname\relax\def\urlprefix{URL }\fi
\providecommand{\eprint}[2][]{\url{#2}}

\bibitem[{{Kennicutt}(1998)}]{kennicutt1998a}
{Kennicutt}, R.~C., Jr. 1998, \apj, 498, 541. \eprint{astro-ph/9712213}

\bibitem[{{K\"oppen} et~al.(1995){K\"oppen}, {Theis}, \&
  {Hensler}}]{koeppen1995a}
{K\"oppen}, J., {Theis}, C., \& {Hensler}, G. 1995, \aap, 296, 99

\bibitem[{{K\"oppen} et~al.(1998){K\"oppen}, {Theis}, \&
  {Hensler}}]{koeppen1998a}
--- 1998, \aap, 331, 524

\bibitem[{{Schmidt}(1959)}]{schmidt1959a}
{Schmidt}, M. 1959, \apj, 129, 243

\end{thebibliography}

\end{document}